\documentclass[11pt,a4paper,american]{article}
\pdfoutput=1
\usepackage{lmodern}

\usepackage[T1]{fontenc}
\usepackage[latin9]{inputenc}
\usepackage{mathrsfs}
\usepackage{amsmath}
\usepackage{amssymb}
\usepackage{esint}

\makeatletter

\pdfpageheight\paperheight
\pdfpagewidth\paperwidth

\usepackage{jcappub}
\numberwithin{equation}{section}

\pdfoutput=1
\renewcommand\[{\begin{equation}}
\renewcommand\]{\end{equation}}

\makeatother

\usepackage{babel}
\begin{document}

\title{Axionic cosmological constant }

\subheader{preprint number }

\author[a]{Katrin Hammer,}

\author[b,c]{Pavel Jirou\v{s}ek,}

\author[b]{and Alexander Vikman}

\affiliation[a]{\em Arnold Sommerfeld Center for Theoretical Physics, }

\affiliation{\em Ludwig Maximilian University Munich, 
Theresienstr. 37, D-80333, Munich, Germany\\
}

\affiliation[b]{\em CEICO-Central European Institute for Cosmology and Fundamental Physics, }

\affiliation{\em Institute of Physics of the Czech Academy of Sciences,\\
Na Slovance 1999/2, 18221 Prague 8, Czech Republic\\
}

\affiliation[c]{\em Institute of Theoretical Physics, Faculty of Mathematics and Physics, Charles University, }

\affiliation{\em V Hole\v{s}ovi\v{c}k\'ach 2, 180 00 Prague 8, Czech Republic}

\emailAdd{vikman@fzu.cz}

\emailAdd{jirousek@fzu.cz}

\emailAdd{katrin.a.hammer@physik.uni-muenchen.de}

\abstract{We propose a novel higher-derivative, Weyl-invariant and generally-covariant
theory for the cosmological constant. This theory is a mimetic construction
with gauge fields playing the role of dynamical variables. These fields
compose Chern-Simons current instead of the vector field in the Henneaux
and Teitelboim formulation of the unimodular gravity. The equations
of motion exactly reproduce the traceless Einstein equations. We demonstrate
that, reformulated in Weyl-invariant variables, this novel theory
reduces to standard general relativity with the cosmological constant
as a Lagrange multiplier. This Lagrange multiplier has an axion-like
coupling. }
\maketitle

\section{Introduction and Discussion }

More than a century ago Einstein proposed \cite{Einstein:1919gv}
the trace-free part of his now standard general relativity (GR) equations,
as a foundation for the dynamics of spacetime. If one assumes energy-momentum
conservation, this formulation is classically equivalent to GR with
a cosmological constant that is now demoted from the status of a constant
of nature to a mere constant of integration, for recent discussion
see e.g. \cite{Weinberg:1988cp,Ellis:2010uc,Ellis:2013eqs}. The value
of this integration constant is fixed by initial conditions. Thus,
the current cosmological constant may be a remnant from the early
universe quantum gravity era, for discussion see e.g. \cite{Linde:1984ir,Linde:2015edk}.
The main advantage of this formulation is that these traceless Einstein
equations are invariant with respect to the vacuum shifts of the matter
energy-momentum tensor 
\[
T_{\mu\nu}\rightarrow T_{\mu\nu}+\Lambda g_{\mu\nu}\,.
\]
This rather desirable property \cite{Padmanabhan:2019art} puts the
cosmological constant problem in a rather different perspective. 

The cosmological constant problem is one of the most important problems
in modern physics, for reviews and recent pedagogical expositions
see e.g. \cite{Weinberg:1988cp,Martin:2012bt,Burgess:2013ara,Padilla:2015aaa}.
The essence of the cosmological constant (CC) problem is a fine-tuning
of the value of the observed acceleration of our expanding universe
caused by this CC. However, any discussion of naturalness and anthropic
reasoning implicitly assumes that the cosmological constant or vacuum
energy can take different values for different solutions. Hence, theories
where CC is not a constant of nature, but a dynamical variable are
useful to apply anthropic reasoning. 

As it was realized already by Einstein one can obtain such a formulation
of gravitational dynamics form the Einstein-Hilbert action where one
does not vary the determinant of the metric. Therefore very often
such theories are referred to as ``unimodular gravity''. There are
different formulations of the dynamics of the unimodular gravity,
see e.g. \cite{Henneaux:1989zc,Buchmuller:1988wx}. The most relevant
for us is the generally-covariant construction by Henneaux and Teitelboim
(HT) \cite{Henneaux:1989zc} with a vector field and our recently
introduced mimetic reformulation \cite{Jirousek:2018ago} based on
a vector field of conformal weight four. 

In the current paper we upgraded our previous Weyl-invariant proposal
of mimetic dark energy \cite{Jirousek:2018ago} with an unusual vector
field of conformal weight four to a mimetic construction with common
abelian and non-abelian /Yang-Mills gauge fields. In the current modification,
the role of this vector field is played by a Chern-Simons current.
Similarly to the original mimetic dark matter proposal \cite{Chamseddine:2013kea},
the theory has higher derivatives, but does not suffer from the Ostrogradsky
ghosts because of the gauge degeneracy caused by the Weyl invariance.
Our mimetic setup allows for a reformulation without higher derivatives.
In this reformulation, the cosmological constant possesses an axion
like coupling envisioned long time ago by Wilczek in \cite{Wilczek:1983as}:
``I would like to briefly mention one idea in this regard, that I
am now exploring. It is to do something for the A-parameter very similar
to what the axion does for the $\theta$-parameter in QCD, another
otherwise mysteriously tiny quantity. The basic idea is to promote
these parameters to dynamical variables, and then see if perhaps small
values will be chosen dynamically.'' However, here we have not touched
any dynamical mechanisms allowing to obtain value of CC corresponding
to the current data. 

The paper is organized in the following way. In section (\ref{sec:Mimetic-Construction-with})
we refresh the relevant elements from \cite{Jirousek:2018ago}, introduce
our mimetic construction for an abelian gauge field and connect this
model to our previous proposal \cite{Jirousek:2018ago}. In section
(\ref{sec:Built-in-constraints}) we discuss identities which appear
per construction in all mimetic theories and in particular in the
theory proposed here. Then, in section (\ref{sec:Matching-Gauge-Transformations?}),
we discuss the correspondence between the U(1) gauge transformations
in the current construction and gauge symmetry corresponding to the
transverse shifts in the vector field in \cite{Jirousek:2018ago}
and \cite{Henneaux:1989zc}. Then we derive the equations of motion
and discuss their properties in section (\ref{sec:Equations-of-motion}).
In section (\ref{sec:Gauge-invariant-variables}) we find Weyl-invariant
variables and reformulate the theory using Lagrange multiplier. Here
we show that the field corresponding to the cosmological constant
has an axion like coupling. After that in section (\ref{sec:Faddeev-Jackiw-Procedure-and})
we follow the Faddeev-Jackiw procedure and analyze the dynamical degrees
of freedom in the abelian case. Here we show that the cosmic time
canonically conjugated to the cosmological constant is given by the
Chern-Simons charge. It is easy to verify that this will also be the
case fo the further non-abelian extensions which we introduce in section
(\ref{sec:Non-abelian-generalization}). There we generalize this
construction to arbitrary SU(N) gauge group and discuss a potential
embedding of this theory as an IR-limit of a confined Yang-Mills theory
(like QCD) with the axion. Finally, in section (\ref{sec:Nonlinear-extension})
we introduce other novel vector-tensor theories describing cosmological
constant as an integration constant. 

Clearly current luck of understanding of the tiny value of the cosmological
constant encourages further investigation of these theoretic constructions. 

\section{Mimetic Construction with Chern-Simons Current\label{sec:Mimetic-Construction-with} }

In the previous work \cite{Jirousek:2018ago} some of us proposed
a new extension of the \emph{mimetic} construction \cite{Chamseddine:2013kea}
incorporating a vector field $V^{\alpha}$. As usual in mimetic theories,
we demoted the spacetime metric, $g_{\mu\nu}$, from its status of
a dynamical variable and introduced the ansatz\footnote{After our previous work \cite{Jirousek:2018ago} was published and
most of the current paper was completed, we became aware that this
ansatz was introduced before in \cite{Kimpton:2012rv} from a completely
different reasoning. We are thankful to Tony Padilla for pointing
out this useful reference. } for $g_{\mu\nu}$:

\begin{equation}
g_{\mu\nu}=h_{\mu\nu}\cdot\left(\nabla_{\alpha}^{\left.h\right)}V^{\alpha}\right)^{1/2}\text{ .}\label{eq:mimetic_vector}
\end{equation}
In this substitution, $g_{\mu\nu}$ \textendash{} is the physical,
free-fall, metric, while $h_{\mu\nu}$ is an \emph{auxiliary} metric
and a new dynamical variable and the covariant derivative, $\nabla_{\alpha}^{\left.h\right)}$,
is the Levi-Civita connection compatible with this 
\begin{equation}
\nabla_{\alpha}^{\left.h\right)}h_{\mu\nu}=0\text{ .}\label{eq:metric_compatibility}
\end{equation}
The main idea behind the ansatz (\ref{eq:mimetic_vector}) was that
any seed theory which originally has $g_{\mu\nu}$ as a dynamical
variable will be mapped into a new \emph{Weyl-invariant} theory with
dynamical variables $h_{\mu\nu}$ and $V^{\mu}$. The latter property
can be realized only if the vector field $V^{\mu}$ has \emph{conformal
weight four} under the Weyl transformations: 
\begin{equation}
V^{\mu}=\Omega^{-4}\left(x\right)V'^{\mu}\,,\qquad\text{along with}\qquad h_{\mu\nu}=\Omega^{2}\left(x\right)h'_{\mu\nu}\,.\label{eq:Weyl_trans_vector_metric}
\end{equation}
On top of the Weyl symmetry above, the resulting theory will possess
another gauge invariance 
\begin{equation}
V^{\mu}=V'^{\mu}+\xi^{\mu}\,,\qquad\text{where}\qquad\nabla_{\mu}^{\left.h\right)}\xi^{\mu}=0\,,\label{eq:degeneracy}
\end{equation}
as the latter preserves the ansatz (\ref{eq:mimetic_vector}). 

As it was demonstrated in the previous work \cite{Jirousek:2018ago},
if the seed theory is general relativity (GR) with the Einstein-Hilbert
action, then the resulting theory, written in Weyl-invariant variables,
reduces to the Henneaux and Teitelboim (HT) generally-covariant formulation
\cite{Henneaux:1989zc} of the so-called ``unimodular gravity''.
Interestingly, this classical equivalence along with the Weyl-invariance
of the resulting theory seem to be overlooked in \cite{Kimpton:2012rv}. 

In the resulting theory there is only one global additional dynamical
degree of freedom (on top of the usual two graviton polarizations)
which is given by the integral\footnote{We use: the standard notation $\sqrt{-h}\equiv\sqrt{-\text{det}h_{\mu\nu}}$
, the signature convention $\left(+,-,-,-\right)$, and the units
$c=\hbar=1$, $M_{\text{Pl}}=\left(8\pi G_{\text{N}}\right)^{-1/2}=1$. } 
\begin{equation}
\mathscr{T}\left(t\right)=\int_{t}d^{3}\mathbf{x}\sqrt{-h}\,V^{t}\left(t,\mathbf{x}\right)\,.\label{eq:dof_HT}
\end{equation}
This dynamical degree of freedom - ``cosmic time'' - is canonically
conjugated to the cosmological constant $\Lambda$, and is shifted
by a \emph{constant }
\begin{equation}
\mathscr{T}\left(t\right)=\mathscr{T}'\left(t\right)+c\,,\label{eq:shift}
\end{equation}
as a result of the gauge transformations (\ref{eq:degeneracy}). As
usual in classical mechanics, shift-symmetry in a coordinate results
in conservation of the canonical momentum: $\Lambda=const$. Here
it is crucial that similarly to current conservation, the divergence-free
condition for the gauge transformations (\ref{eq:degeneracy}), $\nabla_{\mu}^{\left.h\right)}\xi^{\mu}=0$,
results in $\int_{t}d^{3}\mathbf{x}\sqrt{-h}\,\xi^{t}\left(t,\mathbf{x}\right)=const$
which we denoted as $c$. 

Contravariant vector fields of conformal weight four are rather unusual
objects. A common conformal weight for a \emph{contravariant} vector
field is \emph{two}. This suggests to look for more fundamental vector
objects which can compose $V^{\mu}$. 

In this paper we propose a new vector mimetic construction where $V^{\mu}$
is not a dynamical variable, but is composed out of usual gauge fields
$A_{\mu}$. This option was also briefly mentioned in \cite{Kimpton:2012rv},
but again from a completely different perspective. Namely, we propose
the ansatz\footnote{It would be more proper to use the absolute value of the invariant
$F_{\alpha\beta}\widetilde{F}^{\alpha\beta}$ in this and consequent
formulas, but for simplicity we omit the absolute value.}
\begin{equation}
g_{\mu\nu}=h_{\mu\nu}\cdot\sqrt{F_{\alpha\beta}\widetilde{F}^{\alpha\beta}}\,,\label{eq:Metric_Ansatz}
\end{equation}
for the spacetime metric in the Einstein-Hilbert action. In this ansatz
\begin{equation}
F_{\mu\nu}=\partial_{\mu}A_{\nu}-\partial_{\nu}A_{\mu}\,,\label{eq:Field_strength}
\end{equation}
 is the usual field strength tensor for a U(1) gauge potential $A_{\mu}$,
while $\widetilde{F}^{\alpha\beta}$ is the corresponding Hodge-dual
tensor 
\begin{equation}
\widetilde{F}^{\alpha\beta}=\frac{1}{2}\cdot\frac{\epsilon^{\alpha\beta\mu\nu}}{\sqrt{-h}}\cdot F_{\mu\nu}\,,\label{eq:Hodge_dual}
\end{equation}
with the Levi-Civita symbol\footnote{We use the convention $\epsilon^{0123}=+1$. }
$\epsilon^{\alpha\beta\mu\nu}$ and the related tensor $E^{\alpha\beta\mu\nu}=\epsilon^{\alpha\beta\mu\nu}/\sqrt{-h}$.
Later in this paper we generalize this ansatz (\ref{eq:Metric_Ansatz})
to a non-abelian case. One can rewrite the physical metric as 
\begin{equation}
g_{\mu\nu}=\frac{h_{\mu\nu}}{\left(-h\right)^{1/4}}\cdot\sqrt{\mathcal{P}}\,,\label{eq:g_with_P}
\end{equation}
where 
\begin{equation}
\mathcal{P}=\frac{1}{2}\epsilon^{\alpha\beta\mu\nu}F_{\alpha\beta}F_{\mu\nu}=2\epsilon^{\alpha\beta\mu\nu}\partial_{\alpha}A_{\beta}\partial_{\mu}A_{\nu}\,,\label{eq:Chern-Pontryagin}
\end{equation}
is the Chern-Pontryagin density, which is insensitive to the metric.
Hence, in our ansatz, for the determinant of the physical metric we
have
\begin{equation}
g=-\mathcal{P}^{2}\,.\label{eq:detg_P}
\end{equation}

The physical metric $g_{\mu\nu}$ defined through (\ref{eq:Metric_Ansatz})
is invariant under the Weyl transformations of the auxiliary metric
$h_{\mu\nu}$. Indeed, as usual, the gauge field $A_{\mu}$ is invariant
under the Weyl transformations and so it is the case for the Chern-Pontryagin
density, $\mathcal{P}$. Hence, the invariance of $g_{\mu\nu}$ directly
follows from (\ref{eq:g_with_P}). 

Now we can recall that 
\[
F_{\alpha\beta}\widetilde{F}^{\alpha\beta}=E^{\alpha\beta\mu\nu}\left(\nabla_{\alpha}^{\left.h\right)}A_{\beta}\right)F_{\mu\nu}=\nabla_{\alpha}^{\left.h\right)}\left(E^{\alpha\beta\mu\nu}A_{\beta}F_{\mu\nu}\right)\,,
\]
as due to the Bianchi identity $\nabla_{\alpha}^{\left.h\right)}\left(E^{\alpha\beta\mu\nu}F_{\mu\nu}\right)=0$.
Hence, one can introduce the so-called \emph{Chern-Simons current},
$C^{\alpha}$, which is given by 
\begin{equation}
C^{\alpha}=E^{\alpha\beta\mu\nu}A_{\beta}F_{\mu\nu}=2\widetilde{F}^{\alpha\beta}A_{\beta}=2E^{\alpha\beta\mu\nu}A_{\beta}\nabla_{\mu}^{\left.h\right)}A_{\nu}=2E^{\alpha\beta\mu\nu}A_{\beta}\partial_{\mu}A_{\nu}\,,\label{eq:Chern-Simons_current}
\end{equation}
where we listed some of the useful identities. Under the Weyl transformations
this pseudovector has conformal weight four as 
\begin{equation}
C^{\alpha}=2\frac{\epsilon^{\alpha\beta\mu\nu}}{\sqrt{-h}}A_{\beta}\partial_{\mu}A_{\nu}=2\Omega^{-4}\left(x\right)\frac{\epsilon^{\alpha\beta\mu\nu}}{\sqrt{-h'}}A_{\beta}\partial_{\mu}A_{\nu}=\Omega^{-4}\left(x\right)C'^{\alpha}\,.\label{eq:Weyl_transofrm_C}
\end{equation}
We can now write our novel mimetic vector ansatz (\ref{eq:Metric_Ansatz})
as 
\begin{equation}
g_{\mu\nu}=h_{\mu\nu}\cdot\sqrt{\nabla_{\alpha}^{\left.h\right)}C^{\alpha}}\,,\label{eq:VthrougC_Ident}
\end{equation}
and identify the Chern-Simons current (ChS) with the vector field
$V^{\mu}$ introduced in \cite{Jirousek:2018ago}. It is important
to stress that the dynamical variables in the action are $\left\{ A_{\mu},h_{\alpha\beta}\right\} $.
Contrary to ansatz (\ref{eq:mimetic_vector}) from our previous work
\cite{Jirousek:2018ago}, here, the physical metric $g_{\mu\nu}$
does not depend on the derivatives of the auxiliary metric $h_{\alpha\beta}$.
The price for this simplification is that the field $V^{\mu}$ becomes
the ChS current, which is a composite variable quadratic in the elementary
dynamical variables $A_{\mu}$. Moreover, one should be rather vigilant
in using this identification, as a straight field redefinition, $V^{\mu}\rightarrow A_{\mu}$,
because (\ref{eq:Chern-Simons_current}) contains time derivatives
$\partial_{t}A_{\mu}$. These time derivatives would usually imply
that the equations of motion for $\left\{ A_{\mu},h_{\alpha\beta}\right\} $
may have more solutions than the equations of motion for $\left\{ V^{\mu},h_{\alpha\beta}\right\} $.
Thus, the proposed construction should not necessarily result in a
theory equivalent to the one introduced in \cite{Jirousek:2018ago}
corresponding to the ``unimodular gravity''. Nevertheless, we will
show just a bit later that, indeed, both theories, with vector of
conformal weight four (\ref{eq:Metric_Ansatz}) and with gauge field
(\ref{eq:mimetic_vector}) have the cosmological constant as an integration
constant and no other dynamical degrees of freedom that would be extra
to those in standard GR. The key hint for this equivalence is that
the true dynamical degree fo freedom of the original theory (\ref{eq:dof_HT}),
the cosmic time, only depends on $V^{t}$, which after identification
becomes $C^{t}$ given by (\ref{eq:Chern-Simons_current}). However,
the later does not contain dangerous time-derivatives of the gauge
potential $A_{\mu}$. 

Substitution of the ansatz (\ref{eq:mimetic_vector}) into \emph{any}
seed action functional $S\left[g,\Phi_{m}\right]$ (with some matter
fields $\Phi_{m}$) induces a \emph{novel} Weyl-invariant theory with
the action functional 
\begin{equation}
S\left[h,A,\Phi_{m}\right]=S\left[g\left(h,A\right),\Phi_{m}\right]\,.\label{eq:action_transform}
\end{equation}

Now we can plug in the ansatz (\ref{eq:mimetic_vector}) into the
Einstein-Hilbert action to obtain an action for a higher-derivative
and U(1)-invariant vector-tensor theory
\begin{equation}
S_{g}\left[h,A\right]=-\frac{1}{2}\int d^{4}x\sqrt{-h}\left[\left(F_{\alpha\beta}\widetilde{F}^{\alpha\beta}\right)^{1/2}\text{ }R\left(h\right)+\frac{3}{8}\cdot\frac{\left(\nabla_{\mu}^{\left.h\right)}\left(F_{\alpha\beta}\widetilde{F}^{\alpha\beta}\right)\right)^{2}}{\left(F_{\sigma\rho}\widetilde{F}^{\sigma\rho}\right)^{3/2}}\right]\text{ .}\label{eq:ShA}
\end{equation}
From this action it is clear that the invariant $F_{\alpha\beta}\widetilde{F}^{\alpha\beta}$
can never vanish on any physical solution. This is clearly a novel
U(1)-invariant scalar-vector theory going beyond Horndeski and other
more recent constructions. For details see \cite{Heisenberg:2018vsk,Clifton:2011jh}.
The total action is $S\left[h,A,\Phi_{m}\right]=S_{g}\left[h,A\right]+S_{m}\left[h,A,\Phi_{m}\right]$
where the second term is the action for matter fields. It is convenient
to introduce a scalar field $\varphi$
\begin{equation}
F_{\alpha\beta}\widetilde{F}^{\alpha\beta}=\left(\frac{\varphi^{2}}{6}\right)^{2}\,,\label{eq:phi_intro}
\end{equation}
so that all other matter fields are coupled to the physical metric
\begin{equation}
g_{\mu\nu}=\frac{\varphi^{2}}{6}\cdot h_{\mu\nu}\,.\label{eq:physical_metric_phi}
\end{equation}
An elegant representation of the ansatz (\ref{eq:Metric_Ansatz})
comes from the identity 
\[
\text{det}F_{\mu\nu}=-\left(\frac{F_{\alpha\beta}\widetilde{F}^{\alpha\beta}}{4}\right)^{2}\cdot h\,,
\]
implying that
\[
g_{\mu\nu}=2h_{\mu\nu}\cdot\left(-\frac{\text{det}F_{\mu\nu}}{h}\right)^{1/4}\,.
\]

It is important to stress that the Weyl-invariance does not fix the
form of the mimetic ansatz. Indeed, in \cite{Gorji:2018okn,Gorji:2019ttx}
it was proposed to use another invariant of the field strength tensor
so that $g_{\mu\nu}=h_{\mu\nu}\cdot\sqrt{F_{\alpha\beta}F^{\alpha\beta}}$
instead of (\ref{eq:Metric_Ansatz}). This ansatz also preserves the
Weyl-invariance of the metric $g_{\mu\nu}$, but instead of the cosmological
constant the result mimics the spatial curvature, at least in cosmology. 

Interestingly, similarly to these works \cite{Gorji:2018okn,Gorji:2019ttx},
one can show that among conformal substitutions $g_{\mu\nu}=C\left(F_{\alpha\beta}\widetilde{F}^{\alpha\beta}\right)h_{\mu\nu}$
or $g_{\mu\nu}=C\left(\nabla_{\alpha}^{\left.h\right)}V^{\alpha}\right)h_{\mu\nu}$,
the only $C$ that allows for a degeneracy and that induces a new
theory and not just a mere field-redefinition\footnote{For the similar discussion in context of the scalar-field mimetic
construction and more general disformal transformations see \cite{Zumalacarregui:2013pma,Deruelle:2014zza},
while for degeneracy in gauge field metric transformations see \cite{DeFelice:2019hxb,Gumrukcuoglu:2019ebp}.}, is a square root. Thus the degeneracy and the appearance of new
(global) degree of freedom is directly linked to the Weyl symmetry. 

\section{Built-in Constraints\label{sec:Built-in-constraints} }

Usually mimetic theories possess a built-in constraint involving the
physical metric, $g_{\mu\nu}$ and other Weyl-invariant quantities
composed out of the dynamical variables. For instance, the original
mimetic dark matter \cite{Chamseddine:2013kea} has the built-in Hamilton-Jacobi
equation 
\begin{equation}
g^{\mu\nu}\,\partial_{\mu}\phi\,\partial_{\nu}\phi=1\,.\label{eq:Hamilton_Jacobi}
\end{equation}
 While the mimetic dark energy from \cite{Jirousek:2018ago} has 
\begin{equation}
\nabla_{\mu}^{\left.g\right)}W^{\mu}=1\,,\label{eq:build_in_constraint_DE}
\end{equation}
where $\nabla_{\mu}^{\left.g\right)}$ is the Levi-Civita connection
compatible with the physical metric 
\begin{equation}
\nabla_{\mu}^{\left.g\right)}g_{\alpha\beta}=0\text{ ,}\label{eq:metric_compatibility_physical}
\end{equation}
and therefore Weyl-invariant with respect to $h_{\mu\nu}=\Omega^{2}\left(x\right)h'_{\mu\nu}$,
while the Weyl-invariant vector field $W^{\mu}$ was defined as 
\[
W^{\mu}=\frac{V^{\mu}}{\nabla_{\alpha}^{\left.h\right)}V^{\alpha}}\,,
\]
with $h_{\mu\nu}-$compatible Levi-Civita connection $\nabla_{\alpha}^{\left.h\right)}$. 

In the current reincarnation of mimetic dark energy it is useful to
introduce $F^{\star\alpha\beta}$ as the Hodge-dual tensor 
\begin{equation}
F^{\star\alpha\beta}=\frac{1}{2}\cdot\frac{\epsilon^{\alpha\beta\mu\nu}}{\sqrt{-g}}\cdot F_{\mu\nu}\,,\label{eq:physical_Hodge_dual}
\end{equation}
defined with respect to the physical metric, $g_{\mu\nu}$. It is
easy to check that, per construction, the analogue of (\ref{eq:Hamilton_Jacobi})
and (\ref{eq:build_in_constraint_DE}) is
\begin{equation}
F_{\alpha\beta}F^{\star\alpha\beta}=1\,.\label{eq:new_built_in_constraint}
\end{equation}
All these constraints are just identities which also hold off-shell.
Usually, a Weyl-invariant formulation of a mimetic theory boils down
to just standard GR supplemented by these constraints that have to
hold now only on-shell - on equations of motion. In section (\ref{sec:Gauge-invariant-variables})
we will see that this expectation is fulfilled also for the theory
under consideration. 

\section{Matching Gauge Transformations?\label{sec:Matching-Gauge-Transformations?} }

It is useful to recall that the Chern-Simons current is not a gauge
invariant object with respect to U(1) transformations. Indeed, under
the usual U(1) gauge transformation 
\begin{equation}
A_{\mu}=A'_{\mu}+\partial_{\mu}\theta\,,\label{eq:U(1)_transform}
\end{equation}
 the ChS current transforms \emph{inhomogeneously} as 
\begin{equation}
C^{\alpha}=C'^{\alpha}+2\widetilde{F}^{\alpha\beta}\partial_{\beta}\theta\,.\label{eq:ChS_transformation}
\end{equation}
 However, the good news are that 
\[
\nabla_{\alpha}^{\left.h\right)}\left(\widetilde{F}^{\alpha\beta}\partial_{\beta}\theta\right)=0\,,
\]
due to the Bianchi identity, commutativity of derivatives and antisymmetry
of $F_{\mu\nu}$. In this way, it seems that at least some of the
gauge transformations (\ref{eq:degeneracy}) (and consequently the
global shifts of the cosmic time (\ref{eq:shift})) can be generated
by the usual U(1) gauge transformations (\ref{eq:U(1)_transform})
with a particular vector field 
\begin{equation}
\xi_{\left(\theta\right)}^{\alpha}=2\widetilde{F}^{\alpha\beta}\partial_{\beta}\theta\,.\label{eq:map_of_gauges}
\end{equation}
However, not all of the gauge transformations (\ref{eq:degeneracy})
(global shifts of the cosmic time (\ref{eq:shift})) can be represented
through the so-called small U(1) gauge transformations, with the gauge
function vanishing at infinity. Indeed, a general divergence-free
vector field $\xi^{\mu}$ contains three independent functions, whereas
in U(1) transformations there is only one free function $\theta$. 

Moreover, one can show that the global shifts in cosmic time (\ref{eq:shift})
could be only generated by \emph{large} gauge transformations that
have $\theta$ non-vanishing at the spatial boundary of spacetime,
$\mathcal{B}$. Indeed, 
\begin{equation}
\int_{t}d^{3}\mathbf{x}\sqrt{-h}\,\xi_{\left(\theta\right)}^{t}=\int_{t}d^{3}\mathbf{x}\,\epsilon^{tikm}F_{km}\partial_{i}\theta=\int_{t}d^{3}\mathbf{x}\partial_{i}\left(\theta\epsilon^{tikm}F_{km}\right)=\oint_{\mathcal{B}}ds_{i}\,\theta\epsilon^{tikm}F_{km}\,,\label{eq:vanishing_time_shift}
\end{equation}
where we used (\ref{eq:Hodge_dual}), the Bianchi identity and the
3d Stokes theorem. Clearly, this integral can only be nonvanishing
provided the gauge functions $\theta$ and the component of magnetic
field normal to the spatial boundary surface are both nonvanishing.
Thus, it seems that off-shell properties of these two theories are
not identical. 

\section{Equations of Motion\label{sec:Equations-of-motion} }

Let us derive the equations of motion for our novel vector-tensor
theory (\ref{eq:ShA}). For the variation of the total action we get
\begin{equation}
\delta S=\frac{1}{2}\int d^{4}x\sqrt{-g}\left(T_{\mu\nu}-G_{\mu\nu}\right)\delta g^{\mu\nu}+\text{boundary terms}\,,\label{eq:variation_action}
\end{equation}
where $G_{\mu\nu}$ is the Einstein tensor for the \emph{physical}
metric $g_{\mu\nu}$ and the energy momentum tensor of matter is defined
as usual with respect to the physical metric 
\begin{equation}
T_{\mu\nu}=\frac{2}{\sqrt{-g}}\cdot\frac{\delta S_{m}}{\delta g^{\mu\nu}}\,.\label{eq:EMT_standard}
\end{equation}
The variation of the contravariant physical metric gives 
\begin{equation}
\delta g^{\mu\nu}=\frac{6}{\varphi^{2}}\cdot\delta h^{\mu\nu}-2g^{\mu\nu}\cdot\frac{\delta\varphi}{\varphi}\,,\label{eq:variation_metric}
\end{equation}
where the variation of the scalar can be expressed from (\ref{eq:phi_intro})
as 
\begin{equation}
4F_{\alpha\beta}\widetilde{F}^{\alpha\beta}\cdot\frac{\delta\varphi}{\varphi}=\delta\left(F_{\alpha\beta}\widetilde{F}^{\alpha\beta}\right)=4E^{\alpha\beta\mu\nu}\nabla_{\alpha}^{h)}A_{\beta}\nabla_{\mu}^{h)}\delta A_{\nu}+\frac{1}{2}F_{\alpha\beta}\widetilde{F}^{\alpha\beta}h_{\mu\nu}\delta h^{\mu\nu}\,.\label{eq:variation_divergence}
\end{equation}
Then from (\ref{eq:variation_action}) and (\ref{eq:variation_metric}),
using the Bianchi identity, one obtains the equation of motion for
the gauge potential 
\begin{equation}
\frac{1}{\sqrt{-h}}\cdot\frac{\delta S}{\delta A_{\nu}}=\widetilde{F}^{\mu\nu}\partial_{\mu}\left(T-G\right)=0\,,\label{eq:EoM_A}
\end{equation}
where $T=T_{\alpha\beta}g^{\alpha\beta}$ and $G=G_{\alpha\beta}g^{\alpha\beta}$.
The Hodge dual of the field tensor has an inverse provided $F_{\alpha\beta}\widetilde{F}^{\alpha\beta}\neq0$,
for a recent discussion see \cite{DeFelice:2019hxb}. Hence the equation
of motion for the gauge field implies 
\begin{equation}
\partial_{\mu}\left(T-G\right)=0\,.\label{eq:Diff_Traces}
\end{equation}
The equation of motion for the auxiliary metric reads 
\[
\frac{1}{\sqrt{-g}}\cdot\frac{\delta S}{\delta h^{\alpha\beta}}=\frac{3}{\varphi^{2}}\left[T_{\alpha\beta}-G_{\alpha\beta}-\frac{1}{4}\left(T-G\right)g_{\alpha\beta}\right]=0\,.
\]
 Hence the equation of motion for the metric $h_{\mu\nu}$ alone directly
reproduces the trace-free part of the Einstein equations 
\begin{equation}
G_{\alpha\beta}-T_{\alpha\beta}-\frac{1}{4}g_{\alpha\beta}\left(G-T\right)=0\,.\label{eq:trace_free_Einstein}
\end{equation}
We would like to stress again that these equations are clearly symmetric
with respect to the vacuum shifts of the energy-momentum tensor for
the matter fields 
\[
T_{\mu\nu}\rightarrow T_{\mu\nu}+\Lambda g_{\mu\nu}\,,
\]
a property which was considered to be desirable already by Einstein
\cite{Einstein:1919gv} more than a century ago. For more recent discussions
see e.g. \cite{Ellis:2010uc,Ellis:2013eqs,Padmanabhan:2019art}. 

Interestingly, the equation of motion for the gauge field $A_{\mu}$
is a direct consequence of the equations of motion for the tensor
$h_{\mu\nu}$ and the equations of motion of the matter. This looks
like a direct consequence of the second Noether theorem, see e.g.
discussion in \cite{Brading:2000hc}. 

Crucially, both equations of motion (\ref{eq:EoM_A}) and (\ref{eq:trace_free_Einstein})
are second order PDE when written in terms of the manifestly Weyl-invariant
composite metric $g_{\mu\nu}$. However, considered as an equation
of motion for original dynamical variables, (\ref{eq:EoM_A}) has
fourth derivatives of $A_{\mu}$ and third of $h_{\mu\nu}$, while
the trace-free part of the $g-$Einstein equations (\ref{eq:trace_free_Einstein})
has up to third derivatives of these original variables $A_{\mu}$. 

The resulting traceless Einstein equations correspond to those of
the so-called \emph{unimodular }gravity. From the point of view of
classical physics the only difference from standard GR is that the
cosmological constant is an \emph{integration constant}. Indeed, integrating
the equation of motion (\ref{eq:EoM_A}) for the gauge field one obtains
\[
G-T=4\Lambda=\text{const}\,,
\]
which one can substitute into (\ref{eq:trace_free_Einstein}). However,
from the point of view of quantum mechanics the cosmological constant
is now promoted to an operator. Consequently, the observed value of
CC (in generic quantum state of the universe) will have quantum fluctuations,
which are per definition impossible for a fixed constant of nature. 

\section{Gauge Invariant Variables and Scalar-Vector-Tensor Formulation\label{sec:Gauge-invariant-variables} }

Now we can follow a similar procedure as in \cite{Hammer:2015pcx}
and \cite{Jirousek:2018ago} and promote the scalar $\varphi$ to
an independent dynamical variable in order to eliminate the higher
derivatives from the action. On this path we introduce a Lagrange
multiplier, $\lambda$, enforcing the definition (\ref{eq:phi_intro})
so that the action (\ref{eq:ShA}) transforms into 
\begin{equation}
S\left[h,\varphi,A,\lambda\right]=\int d^{4}x\sqrt{-h}\left[-\frac{1}{2}\left(\partial\varphi\right)^{2}-\frac{1}{12}\varphi^{2}\text{ }R\left(h\right)-\frac{\lambda}{72}\varphi^{4}+\frac{\lambda}{2}\cdot F_{\alpha\beta}\widetilde{F}^{\alpha\beta}\right]\,.\label{eq:canonical_normalization}
\end{equation}
Hence, we have rewritten (\ref{eq:ShA}) as a scalar-vector-tensor
theory in this way. This theory should be Weyl-invariant, as it was
the case with the original action (\ref{eq:ShA}). This requirement
forces the Lagrange multiplier $\lambda$ to be invariant under the
Weyl transformations. All other matter fields are coupled to the scalar
$\varphi$ through the physical metric 
\[
g_{\mu\nu}=\frac{\varphi^{2}}{6}\cdot h_{\mu\nu}\,.
\]

As we have already noticed in our previous work \cite{Jirousek:2018ago},
the first three terms correspond to the Dirac theory of the Weyl-invariant
gravity \cite{Dirac:1973gk}, see also \cite{Deser:1970hs}. These
terms are also used in the so-called Conformal Inflation \cite{Kallosh:2013hoa}.
The auxiliary scalar field $\varphi$ has a ghost-like kinetic term.
The would be coupling constant $\lambda$ is actually a Lagrange multiplier
field which has a crucial axion-like coupling. 

The form of the action is also closely related to those theories studied
in \cite{Alvarez:2006uu} which are, however, only invariant with
respect to transverse diffeomorphisms preserving the value of the
metric determinant. 

The form of the action suggests that the Weyl symmetry in this setup
is in a sense empty (or as sometimes called fake or sham) and does
not have any dynamical consequence, see e.g. \cite{Tsamis:1984hh,Jackiw:2014koa,Oda:2016pok}. 

The dynamical variables $\left\{ h_{\mu\nu},A_{\mu},\lambda,\varphi\right\} $
transform as 
\begin{align}
 & h_{\mu\nu}=\Omega^{2}\left(x\right)h'_{\mu\nu}\,,\label{eq:Weyl_Transform_set}\\
 & \varphi=\Omega^{-1}\left(x\right)\varphi'\,,\nonumber \\
 & A_{\mu}=A'_{\mu}\,,\nonumber \\
 & \lambda=\lambda'\text{ .}\nonumber 
\end{align}
Instead, of these dynamical variables one can introduce a new set
of independent dynamical variables $\left\{ g_{\mu\nu},A_{\mu},\Lambda,\varphi\right\} $,
where 
\begin{align}
 & g_{\mu\nu}=\frac{\varphi^{2}}{6}\cdot h_{\mu\nu}\,,\label{eq:gauge_invariant_variables}\\
 & \Lambda=\frac{\lambda}{2}\text{ ,}\nonumber 
\end{align}
are gauge invariant. These nonsingular field-redefinitions resemble
the Weyl transformations with $\Omega^{2}=\varphi^{2}/6$, except
we do not reduce the dimensionality of the phase space, as $\varphi$
is not affected by this field-redefinition. 

Performing this field redefinition in (\ref{eq:canonical_normalization})
one rewrites the action as 
\begin{equation}
S\left[g,A,\Lambda,\Phi_{m}\right]=\int d^{4}x\sqrt{-g}\left[-\frac{1}{2}R\left(g\right)+\Lambda\left(F_{\alpha\beta}F^{\star\alpha\beta}-1\right)\right]+S_{m}\left[g,\Phi_{m}\right]\text{ ,}\label{eq:axion_action}
\end{equation}
where the matter is now minimally coupled to gravity and the Hodge
dual $F^{\star\alpha\beta}$ is now defined with respect to the physical
metric $g_{\mu\nu}$, see (\ref{eq:physical_Hodge_dual}). This action
functional does not depend anymore on the conformal factor $\varphi$,
but only on Weyl-invariant dynamical variables $\left\{ g_{\mu\nu},A_{\mu},\Lambda\right\} $.
Clearly the Lagrange multiplier $\Lambda$ has an axion-like coupling.
However, contrary to a normal axion there is no kinetic term. 

The corresponding equations of motion are: the constraint 
\begin{equation}
F_{\alpha\beta}F^{\star\alpha\beta}=1\,,\label{eq:constraint}
\end{equation}
which for non-Weyl-invariant dynamical variables was just built-in
off-shell, see (\ref{eq:new_built_in_constraint}) in section (\ref{sec:Built-in-constraints}),
while the variation with respect to the gauge field yields 
\begin{equation}
\frac{1}{\sqrt{-g}}\cdot\frac{\delta S}{\delta A_{\gamma}}=-4\nabla_{\mu}^{g)}\left(\Lambda E^{\alpha\beta\mu\gamma}\nabla_{\alpha}^{g)}A_{\beta}\right)=4F^{\star\gamma\mu}\partial_{\mu}\Lambda\,,\label{eq:constraint_gauge_inv}
\end{equation}
and the Einstein equations are 
\begin{equation}
\frac{2}{\sqrt{-g}}\cdot\frac{\delta S}{\delta g^{\alpha\beta}}=T_{\alpha\beta}+\Lambda g_{\alpha\beta}-G_{\alpha\beta}=0\,.\label{eq:Einstein_Eq_invar}
\end{equation}
The constraint (\ref{eq:constraint}) enforces that $\widetilde{F}^{\gamma\mu}$
is invertible so that $\Lambda=const$ which also follows from the
Bianchi identity and the conservation of the total energy-momentum.
Clearly these equations are equivalent to those obtained from the
original action, (\ref{eq:EoM_A}) and (\ref{eq:trace_free_Einstein}). 

One possible interpretation of the action (\ref{eq:axion_action}),
as well as the action of the Henneaux-Teitelboim generally-covariant
formulation \cite{Henneaux:1989zc} of the so-called ``unimodular
gravity'' is that they simultaneously describe \emph{all }de Sitter\emph{
}and\emph{ }anti de Sitter\emph{ }universes\emph{. }Interestingly,
the dynamics of these systems is essentially mechanical and not field-theoretical,
as the degrees of freedom are \emph{global }and correspond to the
cosmic time and the canonically conjugated cosmological constant. 

\section{Faddeev-Jackiw Procedure and Degrees of Freedom\label{sec:Faddeev-Jackiw-Procedure-and} }

A proper Hamiltonian analysis of the system would take too much space
in this short paper, and will be reported elsewhere. Indeed, via the
Dirac procedure this analysis can be quite involved and deserved a
separate study in cases of the mimetic gravity e.g. \cite{Ganz:2018mqi}
and of the unimodular gravity e.g. \cite{Kluson:2014esa}. Therefore,
here we will only provide a hint for the structure of the canonical
degrees of freedom in this theory written in the Weyl-invariant variables
(\ref{eq:axion_action}). For this simplified analysis we will follow
the Faddeev-Jackiw procedure \cite{Faddeev:1988qp,Jackiw:1993in}.
It is sufficient to consider the ``constraint part of the action''
(\ref{eq:axion_action}) only. The corresponding Lagrangian density
takes the form 
\[
\mathscr{L}=-\sqrt{-g}\Lambda+2\Lambda\partial_{0}\left(\epsilon^{0ikm}A_{i}\partial_{k}A_{m}\right)+2\Lambda\partial_{k}\left(\epsilon^{k\beta\mu\nu}A_{\beta}\partial_{\mu}A_{\nu}\right)\,,
\]
where the latin indices are purely spatial and run from 1 to 3. Furthermore,
we can introduce a decomposition 
\begin{equation}
A_{i}=\partial_{i}\chi+A_{i}^{T}\,,\qquad\text{where}\qquad\partial_{i}A_{i}^{T}=0\,,\label{eq:decomposition}
\end{equation}
so that a U(1) gauge transformation leaves $A_{i}^{T}$ invariant
and shifts $\chi$ as 
\[
\chi=\chi'+\theta\,,
\]
so that another invariant object is $A_{0}-\dot{\chi}$. 

It is useful to introduce a ``magnetic field'' 
\begin{equation}
B^{i}=\epsilon^{0ikm}\partial_{k}A_{m}^{T}\,.\label{eq:magnetic_field}
\end{equation}
Substituting the decomposition into the Lagrangian, after some algebra,
we obtain 
\[
\mathscr{L}=-\sqrt{-g}\Lambda+2\Lambda\left[\partial_{0}\left(B^{i}A_{i}^{T}\right)+2B^{i}\partial_{i}\left(\dot{\chi}-A_{0}\right)+B^{i}\dot{A}_{i}^{T}-A_{i}^{T}\dot{B}^{i}\right]\,.
\]
In this Lagrangian $\chi$ and $A_{0}$ only enter as a gauge invariant
combination $\dot{\chi}-A_{0}$. Similarly to standard electrodynamics,
$A_{0}$ enforces a constraint while a variation with respect to $\chi$
is superfluous, as it only enforces the time derivative of this constraint
to vanish. Notably, the time derivatives enter this action only linearly,
for this reason this action should be considered as a Hamiltonian
action. Integrating the last two terms by parts and neglecting the
boundary terms one obtains 
\begin{equation}
\int d^{3}\mathbf{x}\Lambda\left(B^{i}\dot{A}_{i}^{T}-A_{i}^{T}\dot{B}^{i}\right)=-\int d^{3}\mathbf{x}\epsilon^{0ikm}\dot{A}_{i}^{T}A_{m}^{T}\partial_{k}\Lambda\,.\label{eq:second_part}
\end{equation}
Covariant equations of motion for the gauge potential (\ref{eq:constraint_gauge_inv})
imply the constraint that the vacuum energy field is constant in space:
$\partial_{i}\Lambda=0$. Hence, if we follow the Faddeev-Jackiw procedure
and plug in this constraint into the action we see that (\ref{eq:second_part})
is vanishing. Moreover, this implies that in the action one can move
the global degree of freedom $\Lambda\left(t\right)$ out of the spatial
integration 
\[
S=\int dtd^{3}\mathbf{x}\mathscr{L}=\int dt\,2\Lambda\int d^{3}\mathbf{x}\,\left[\partial_{0}\left(B^{i}A_{i}^{T}\right)+2B^{i}\partial_{i}\left(\dot{\chi}-A_{0}\right)-\frac{1}{2}\sqrt{-g}\right]\,.
\]
Comparing this result with the usual Hamiltonian form of the action
\[
S=\int dt\left(p\dot{q}-H\left(p,q\right)\right)\,,
\]
provides a hint (if not a strict proof) that $\Lambda$ is canonically
conjugated to the Chern-Simons charge
\[
2\int d^{3}\mathbf{x}\,B^{i}A_{i}^{T}=\int d^{3}\mathbf{x}\sqrt{-g}C_{T}^{t}\,,
\]
where $C_{T}^{t}$ is the U(1)-invariant time component of the of
the Chern-Simons current, which is now constructed out of gauge-invariant
$A_{i}^{T}$ instead of $A_{i}$. As we have already discussed in
(\ref{sec:Matching-Gauge-Transformations?}) small U(1) transformations
do not change the value of the Chern-Simons charge, see (\ref{eq:vanishing_time_shift}).
Thus the cosmic time is given by the Chern-Simons charge which is
a quantity canonically conjugated to the cosmological constant. This
resembles a construction from \cite{Smolin:1994qb,Alexander:2018djy,Magueijo:2018jzg},
where the Chern-Simons charge also appeared as time, but it was composed
out of the imaginary part of the complex Ashtekar connection. Instead
here we use a real connection $A_{\mu}$ which is a four covariant
vector that we attribute to the matter sector. 

\section{Non-abelian Generalization\label{sec:Non-abelian-generalization}}

It is very easy to generalize this theory to any non-abelian gauge
symmetry with an SU(N) gauge group. The motivation for this generalization
is threefold. First, the very structure of vacuum in the non-abelian
theories can be highly involved in IR and due to confinement can possess
nontrivial global degrees of freedom, see e.g. recent discussions
in \cite{Dvali:2005an,Dvali:2005ws,Dvali:2005zk}. On the other hand,
axionic couplings like the one in (\ref{eq:axion_action}) are more
typical for non-abelian field theories. And finally, contrary to the
original system \cite{Jirousek:2018ago} with the unusual vector field
$V^{\mu}$, here one can easily introduce standard couplings of the
otherwise completely auxiliary gauge field $A_{\mu}$ to matter. In
that case the structure of the theory will change, though in some
regimes it can still approximate the cosmological constant. However,
with the coupling to matter the gauge field configurations become
relevant and have to respect the cosmological principle. Yet, it would
not be possible to satisfy homogeneity and isotropy in case of just
one abelian field, with nonvanishing magnetic and electric fields.
The way out could be either to average over many fields like e.g.
in vector inflation \cite{Golovnev:2008cf} or to introduce a triplet
of mutually orthogonal vector fields like in \cite{ArmendarizPicon:2004pm,Bento:1992wy},
or to invoke a non-abelian field theory, as in \cite{Galtsov:1991un,Hosotani:1984wj}.
The later was recently applied in a different mimetic construction
in \cite{Babichev:2017lrx,Babichev:2018twg,Tolley:2009fg}. Especially
interesting is the SU(2) case, as there three independent components
of the gauge potential can play the role of the aforementioned triplet. 

As it is common in the Yang-Mills theories, the non-abelian field
can be expanded into generators $\boldsymbol{A}_{\mu}=A_{\mu}^{c}\boldsymbol{T}_{c}$,
and the covariant derivative $\boldsymbol{D}_{\mu}=\partial_{\mu}+i\text{g}\boldsymbol{A}_{\mu}$
with the self-coupling constant $\text{g}$ yields a curvature 
\[
\boldsymbol{F}_{\mu\nu}=\boldsymbol{D}_{\mu}\boldsymbol{A}_{\nu}-\boldsymbol{D}_{\nu}\boldsymbol{A}_{\mu}=\partial_{\mu}\boldsymbol{A}_{\nu}-\partial_{\nu}\boldsymbol{A}_{\mu}+i\text{g}\left[\boldsymbol{A}_{\mu},\boldsymbol{A}_{\nu}\right]\,,
\]
with the corresponding Hodge dual given by 
\[
\widetilde{\boldsymbol{F}}^{\mu\nu}=\frac{1}{2}E^{\mu\nu\alpha\beta}\boldsymbol{F}_{\alpha\beta}\,.
\]
The ansatz (\ref{eq:Metric_Ansatz}) is then generalized to 
\begin{equation}
g_{\mu\nu}=h_{\mu\nu}\cdot\sqrt{\text{Tr}\boldsymbol{F}_{\alpha\beta}\widetilde{\boldsymbol{F}}^{\alpha\beta}}\,.\label{eq:ansatz_nonabel}
\end{equation}
The corresponding Chern-Simons current is 
\begin{equation}
C^{\mu}=\text{Tr}\,E^{\mu\alpha\beta\gamma}\left(\boldsymbol{F}_{\alpha\beta}\boldsymbol{A}_{\gamma}-\frac{2i\text{g}}{3}\boldsymbol{A}_{\alpha}\boldsymbol{A}_{\beta}\boldsymbol{A}_{\gamma}\right)\,,\label{eq:CS_Nonabelian}
\end{equation}
so that (\ref{eq:VthrougC_Ident}) remains valid. The most important
formulas (\ref{eq:canonical_normalization}) and (\ref{eq:axion_action})
can be generalized by a simple substitution $F_{\alpha\beta}\widetilde{F}^{\alpha\beta}\rightarrow\text{Tr}\boldsymbol{F}_{\alpha\beta}\widetilde{\boldsymbol{F}}^{\alpha\beta}$
(or $\text{Tr}\boldsymbol{F}_{\alpha\beta}\boldsymbol{F}^{\star\alpha\beta}$
when going to Weyl-invariant variables), in particular
\begin{equation}
S\left[g,\boldsymbol{A},\Lambda,\Phi_{m}\right]=\int d^{4}x\sqrt{-g}\left[-\frac{1}{2}R\left(g\right)+\Lambda\left(\text{Tr}\boldsymbol{F}_{\alpha\beta}\boldsymbol{F}^{\star\alpha\beta}-1\right)\right]+S_{m}\left[g,\Phi_{m}\right]\,.\label{eq:Action_nonabel}
\end{equation}
Clearly, the invariant $\sqrt{-g}\,\text{Tr}\boldsymbol{F}_{\alpha\beta}\boldsymbol{F}^{\star\alpha\beta}$
is metric-independent so that the Einstein equations remain (\ref{eq:Einstein_Eq_invar}).
As a trivial consequence of the Bianchi identity and the covariant
conservation of the energy-momentum one obtains vacuum energy as an
integration constant $\Lambda=const$. 

This action literally realizes the axionic coupling of the cosmological
constant envisioned long time ago by Wilczek in \cite{Wilczek:1983as}:
``I would like to briefly mention one idea in this regard, that I
am now exploring. It is to do something for the A-parameter very similar
to what the axion does for the $\theta$-parameter in QCD, another
otherwise mysteriously tiny quantity. The basic idea is to promote
these parameters to dynamical variables, and then see if perhaps small
values will be chosen dynamically.'' However, here we do not discuss
any dynamical selection mechanism. 

There are two important differences of the non-abelian construction
comparing to the abelian case described before. First of all, in the
later, a nonvanishing $F_{\alpha\beta}\widetilde{F}^{\alpha\beta}$
implies the Lorentz symmetry breaking, with both electric and magnetic
field nonvanishing and not orthogonal to each other, whereas in the
non-abelian case a nonvanishing $\text{Tr}\boldsymbol{F}_{\alpha\beta}\boldsymbol{F}^{\star\alpha\beta}$
does not necessarily lead to any Lorentz violation. 

One can speculate that the growing cosmic time can correspond to a
growing winding number or growing complexity of the vacuum. 

Another important distinction is that for the abelian field, there
was no dimensional parameter entering the construction. However, here
after a proper normalization, $\boldsymbol{F}_{\mu\nu}$ is dimensionless
so that $\boldsymbol{A}_{\mu}$ has dimensions of length, and the
coupling constant has dimensions $\text{g}=M^{2}$. Otherwise, one
can canonically normalize the gauge fields and have the dimensionless
coupling constant, however, in that case, one has to divide the expression
under the square root in the ansatz (\ref{eq:ansatz_nonabel}) by
the same $M^{4}$. Hence, an appearance of a mass scale $M$ is unavoidable. 

Another intriguing conjecture is that action (\ref{eq:Action_nonabel})
can appear from a strongly coupled gauge field and a strongly coupled
axion $\Lambda$ with a potential
\begin{equation}
V\left(\Lambda\right)=M^{4}\sin\left(\Lambda/M^{4}\right)=\Lambda-\frac{1}{6}M^{4}\left(\frac{\Lambda}{M^{4}}\right)^{3}+...\,,\label{eq:axion_potential}
\end{equation}
 expanded not around a minimum, but around zero under the assumption
that $M\gg\Lambda$. In this case, the first term in the expansion
generates the constraint (\ref{eq:constraint}). On the other hand
the strong coupling does not allow $\boldsymbol{A}_{\mu}$ and $\Lambda$
to have a wave-like propagation. This suggests that, in the strong-coupling
regime, one can neglect both usual kinetic terms $\text{Tr}\boldsymbol{F}_{\alpha\beta}\boldsymbol{F}^{\alpha\beta}$
and $\left(\partial\Lambda\right)^{2}$. The latter is the main condition
for this approximation. This is somehow reminiscent to the dynamical
regime which happens in completions of mimetic dark matter and k-essence
/$P(X)$ by a globally-charged U(1) scalar field, see e.g. \cite{Babichev:2017lrx,Babichev:2018twg,Tolley:2009fg,Son:2000ht,Son:2002zn}.
There one can also neglect the kinetic term of the radial field. In
the next section we pursue this analogy a bit deeper. 

\section{Nonlinear Extension\label{sec:Nonlinear-extension} }

Of course, it is unusual to expand the potential (\ref{eq:axion_potential})
away from its minimum. Therefore, a natural question is whether one
can neglect $\text{Tr}\boldsymbol{F}_{\alpha\beta}\boldsymbol{F}^{\alpha\beta}$
and $\left(\partial\Lambda\right)^{2}$ on configurations around a
minimum of the potential\footnote{For this discussion, the form of the potential is not important and
should not necessarily be (\ref{eq:axion_potential}). It may be even
not periodic anymore. } $V\left(\Lambda\right)$ or maybe at some other point with nonvanishing
$\Lambda$. Suppose this is the case. In that case the action will
take the form 
\begin{equation}
S\left[g,\boldsymbol{A},\Lambda\right]=\int d^{4}x\sqrt{-g}\left[-\frac{1}{2}R\left(g\right)+\Lambda\,\text{Tr}\boldsymbol{F}_{\alpha\beta}\boldsymbol{F}^{\star\alpha\beta}-V\left(\Lambda\right)\right]\,.\label{eq:action_auxiliarly_field}
\end{equation}
In the action (\ref{eq:action_auxiliarly_field}) the term $\sqrt{-g}\,\text{Tr}\boldsymbol{F}_{\alpha\beta}\boldsymbol{F}^{\star\alpha\beta}$
is still metric-independent and does not contribute to the Einstein
equations, while the last term adds to those only the CC contribution
\begin{equation}
T_{\mu\nu}=V\left(\Lambda\right)g_{\mu\nu}\,,\label{eq:T_for_V(L)}
\end{equation}
which implies that $\partial_{\mu}\Lambda=0$, because of 

For the theory (\ref{eq:action_auxiliarly_field}), the axion field
$\Lambda$ is not a Lagrange multiplier anymore, but rather an auxiliary
field. This implies that one can integrate it out using the relation
\begin{equation}
\text{Tr}\boldsymbol{F}_{\alpha\beta}\boldsymbol{F}^{\star\alpha\beta}=V'\,,\label{eq:L(CS)}
\end{equation}
provided $V''\neq0$, where prime denotes a derivative with respect
to $\Lambda$. In that case we reduce the number of dynamical variables
and obtain 
\begin{equation}
S\left[g,\boldsymbol{A}\right]=\int d^{4}x\sqrt{-g}\left[-\frac{1}{2}R\left(g\right)+f\left(\text{Tr}\boldsymbol{F}_{\alpha\beta}\boldsymbol{F}^{\star\alpha\beta}\right)\right]\,,\label{eq:action_Nonlinear}
\end{equation}
where the function $f$ is given by a Legendre transformation 
\begin{equation}
f\left(\text{Tr}\boldsymbol{F}_{\alpha\beta}\boldsymbol{F}^{\star\alpha\beta}\right)=\Lambda V'-V\,,\label{eq:Legendre}
\end{equation}
where $\Lambda=\Lambda\left(\text{Tr}\boldsymbol{F}_{\alpha\beta}\boldsymbol{F}^{\star\alpha\beta}\right)$
is a solution of (\ref{eq:L(CS)}). On the other hand, one could take
(\ref{eq:action_Nonlinear}) with an arbitrary function $f$ as a
starting point. In that case, the corresponding cosmological constant
term would be still equal to (\ref{eq:T_for_V(L)}), but written as
\[
T_{\mu\nu}=\left(\text{Tr}\boldsymbol{F}_{\alpha\beta}\boldsymbol{F}^{\star\alpha\beta}\,f'-f\right)g_{\mu\nu}\,.
\]
Following our analogy between $W^{\mu}$ and the Chern-Simons current
one can also write 
\begin{equation}
S\left[g,W\right]=\int d^{4}x\sqrt{-g}\left[-\frac{1}{2}R\left(g\right)+f\left(\nabla_{\mu}^{\left.g\right)}W^{\mu}\right)\right]\,,\label{eq:Nonlinear_HT}
\end{equation}
instead of 
\begin{equation}
S\left[g,W,\Lambda\right]=\int d^{4}x\sqrt{-g}\left[-\frac{1}{2}R\left(g\right)+\Lambda\nabla_{\mu}^{\left.g\right)}W^{\mu}-V\left(\Lambda\right)\right]\,,\label{eq:Modified_HT}
\end{equation}
and check that (on equations of motion for $W^{\mu}$) both these
actions also reproduce a cosmological constant term 
\[
T_{\mu\nu}=\left(\nabla_{\mu}^{\left.g\right)}W^{\mu}\,f'-f\right)g_{\mu\nu}=V\left(\Lambda\right)g_{\mu\nu}\,.
\]
Clearly, for $V\left(\Lambda\right)=\Lambda$, the action (\ref{eq:Modified_HT})
corresponds to the original Henneaux and Teitelboim \cite{Henneaux:1989zc}
construction for the ``unimodular gravity''. The equations of motion
both these systems imply $\nabla_{\mu}^{\left.g\right)}W^{\mu}=const$,
which is up to a trivial rescaling reproduce the constraint equation
in the original HT \cite{Henneaux:1989zc} action. Interestingly,
a similar formulation of the ``unimodular gravity'' that was slightly
more complicated than just (\ref{eq:Nonlinear_HT}) and (\ref{eq:Modified_HT}),
was given in \cite{Padilla:2014yea}. 

Thus all these actions: the original Henneaux and Teitelboim \cite{Henneaux:1989zc}
and (\ref{eq:Modified_HT}), (\ref{eq:Nonlinear_HT}), (\ref{eq:action_Nonlinear}),
(\ref{eq:action_auxiliarly_field}) and (\ref{eq:Action_nonabel})
along with our Weyl-invariant constructions (\ref{eq:ShA}) and \cite{Jirousek:2018ago}
describe the cosmological constant as an integration constant which
is also their global degree of freedom. A new feature that the formulations
(\ref{eq:Modified_HT}), (\ref{eq:Nonlinear_HT}), (\ref{eq:action_Nonlinear}),
(\ref{eq:action_auxiliarly_field}) with free functions $f$ or $V$
can achieve is that these functions can be bounded in some range.
This boundedness of $V$or $f$ would imply that, for arbitrary values
of $\Lambda$ (or $\nabla_{\mu}^{\left.g\right)}W^{\mu}$ or $\text{Tr}\boldsymbol{F}_{\alpha\beta}\boldsymbol{F}^{\star\alpha\beta}$),
the resulting cosmological constant is limited to be not arbitrary,
but to lie in some range. This feature is very interesting for theories
where $\Lambda$ (or $\nabla_{\mu}^{\left.g\right)}W^{\mu}$ or $\text{Tr}\boldsymbol{F}_{\alpha\beta}\boldsymbol{F}^{\star\alpha\beta}$)
are stochastic variables. 

It is also worthwhile comparing the above transition from the Lagrange
multiplier formulation to the auxiliary field with the similar transition
in mimetic dark matter and k-essence, see e.g. \cite{Babichev:2017lrx,Babichev:2018twg,Tolley:2009fg,Son:2000ht,Son:2002zn}.
There the theory with Lagrangian density 
\[
\mathscr{L}=\lambda\left(\left(\partial\phi\right)^{2}-1\right)\,,
\]
corresponds to a fluid-like dust with an identically vanishing sound
speed, whereas 
\[
\mathscr{L}=\lambda\left(\partial\phi\right)^{2}-V\left(\lambda\right)\,,
\]
represents k-essence with a nonvanishing speed of sound. Thus, there
the promotion of the Lagrange multiplier to an auxiliary field substantially
changes the theory. Notably, the fluid-like dust is just a singular
case, which in the completion \cite{Babichev:2017lrx,Babichev:2018twg,Tolley:2009fg,Son:2000ht,Son:2002zn}
is only realized under an extreme fine-tuning of identically vanishing
self-interaction. 

\medskip{}

\acknowledgments It is a pleasure to thank Gia Dvali, Ond\v{r}ej
Hul\'{i}k, Andrei Linde, Shinji Mukohyama, Atsushi Naruko, Tony Padilla,
Ippocratis Saltas, Misao Sasaki, Sergei Sibiryakov, Arkady Vainshtein,
Masahide Yamaguchi and Tom Z\l{}o\'{s}nik for useful discussions.
During the final stages of this project, P. J. and A.V. enjoyed a
very warm hospitality of the cosmology group at the Tokyo Institute
of Technology. This productive visit was possible thanks to the JSPS
Invitational Fellowships for Research in Japan (Fellowship ID:S19062)
received by A.V. During last polishing of the current text, A.V. was
delighted in his participation in the KITP program ``From Inflation
to the Hot Big Bang'', so that this research was supported in part
by the National Science Foundation under Grant No. NSF PHY-1748958.
A.V. acknowledges support from the J. E. Purkyn\v{e} Fellowship of
the Czech Academy of Sciences. \\
The work of P. J. and A.V. was supported by the funds from the European
Regional Development Fund and the Czech Ministry of Education, Youth
and Sports (M\v{S}MT): Project CoGraDS - CZ.02.1.01/0.0/0.0/15\_003/0000437.
\\

\bibliographystyle{utphys}
\addcontentsline{toc}{section}{\refname}\bibliography{Unimod}

\end{document}